\documentclass[preprint,showpacs,preprintnumbers,amsmath,amssymb]{revtex4}

\usepackage{graphicx}
\usepackage{dcolumn}
\usepackage{bm}
\usepackage{subfigure}
\usepackage{theorem}

\def\N{{\bf N}}

\def\n{{\bf n}}
\def\P{{\bf P}}
\def\E{{\bf E}}
\def\1{{\bf 1}}
\def\Cov{{\rm Cov}}
\def\I{{\bf I}}
\def\J{{\bf J}}
\def\K{{\bf K}}

\begin{document}

\preprint{}

\title{Velocity correlations of a discrete-time 
totally asymmetric simple-exclusion process
in stationary state on a circle}

\author{Yasuyuki Yamada}
\email{yasuyuki@phys.chuo-u.ac.jp}
\author{Makoto Katori}
\email{katori@phys.chuo-u.ac.jp}
\affiliation{%
Department of Physics,
Faculty of Science and Engineering,
Chuo University, 
Kasuga, Bunkyo-ku, Tokyo 112-8551, Japan 
}%

\date{11 October 2011}

\begin{abstract}
The discrete-time version of totally asymmetric simple-exclusion
process (TASEP) on a finite one-dimensional lattice is studied
with the periodic boundary condition.
Each particle at a site hops to the next site
with probability $0 \leq p \leq 1$,
if the next site is empty.
This condition can be rephrased by the condition
that the number $n$ of vacant sites between the particle
and the next particle is positive.
Then the average velocity is given by a product of
the hopping probability $p$ and
the probability that $n \geq 1$.
By mapping the TASEP to another driven diffusive system called
the zero-range process, it is proved that 
the distribution function of vacant sites in the stationary state
is exactly given by a factorized form.
We define $k$-particle velocity correlation function
as the expectation value of a product of velocities of $k$
particles in the stationary distribution.
It is shown that it does not depend on positions
of $k$ particles on a circle but depends only on the number $k$.
We give explicit expressions for all velocity
correlation functions using the Gauss hypergeometric
functions.
Covariance of velocities of two particles is studied in detail
and we show that velocities become independent asymptotically
in the thermodynamic limit.

\end{abstract}


\maketitle

\section{Introduction}

The one-dimensional {\it totally asymmetric 
simple-exclusion process} (TASEP)
is a minimal statistical-mechanics model
for driven diffusive systems of many particles
with hardcore exclusive interaction
\cite{Lig85,Spo91,SZ95,Lig99,Sch00,MKL09}.
In the present paper, we consider a discrete-time
version of the TASEP on a circle,
{\it i.e.} a one-dimensional finite lattice
with periodic boundary condition,
in which the parallel update rule is applied
\cite{RSSS98}.
This process can be exactly mapped to another driven
diffusive particle system called the
{\it zero-range process} (ZRP),
by regarding each number of vacant sites
between successive particles in the TASEP
as a number of particles at each site
in the ZRP \cite{Eva00,EH05}.

In the ZRP representation, particles hop from
site to site on a lattice with a hopping probability
which depends only on the number of particles
at the departure site.
We assume that there do not occur any creation and annihilation of
particles in the ZRP.
Then, if a site is vacant, there is no possibility
that a particle hops
from that empty site to other sites, as a matter of course.
The configuration in the TASEP such that
the next site of a particle is occupied by
another particle is represented by the configuration
in the ZRP such that the corresponding site is empty.
Therefore the prohibition of hopping 
by hardcore exclusion in such a jamming situation in the TASEP 
is automatically satisfied in the ZRP.
Moreover, the steady state of the ZRP is exactly
described by the probability density function
in a factorized form \cite{Eva00,EH05},
and then the stationary distribution of vacant sites
in the TASEP on a circle is explicitly determined
as Eq.(\ref{eqn:PLN}) given below
\cite{Eva00,EH05,KNT06,Kan07}.

The map from the TASEP to the ZRP is
interesting from the viewpoint of quantum statistical
mechanics, since it seems to be a map from
an interacting Fermi gas to a free Bose gas.
On the other hand, as demonstrated in the present
paper, the procedure in which we analyze the TASEP/ZRP
is quite different from the standard way
for free Boson systems in the following sense.
(i) Thermal equilibrium state of free Bose gas is usually
treated in the grand canonical ensemble by introducing
fugacity, while here we want to consider the
driven diffusive system with a fixed number of particles
and thus we treat the system in the canonical ensemble.
Then the canonical partition function $Z_{L,N}$,
where $L$ and $N$ denote the numbers of
particles and vacancies in the TASEP, respectively, 
plays an important role in calculation.
(ii) In the usual theory of free Bose gases,
{\it condensation} of particles in a specified energy level
is carefully studied ({\it e.g.}, the Bose-Einstein condensation
in a ground state). 
In the context of ZRP, however, distribution of {\it vacancies}
({\it i.e.}, empty energy levels) should be
well studied.
The reason is that,
if a site in the ZRP is occupied by one or
more than one particle, then the velocity
of the corresponding particle in the TASEP
can be positive, but if the site is empty
in the ZRP, the velocity of the particle
in the TASEP is definitely zero.
When we consider the TASEP as a simple model
of traffic flow,
the flux which is defined as the product
of velocity and particle-density is the most
important quantity.
The purpose of the present paper is to study
velocity correlations of particles 
in the stationary state of the TASEP on a circle.

Let $\N=\{1,2,3, \dots\}$ and
$\N_0 =\{0\} \cup \N=\{0,1,2, \dots\}$.
For $K \in \N$, we consider a one-dimensional lattice
$\Lambda=\{1,2, \dots, K \}$.
Each site $i \in \Lambda$ is either occupied by a particle,
which is denoted by $\eta(i)=1$, or vacant denoted by
$\eta(i)=0$.
The following discrete-time stochastic process is considered 
for simulating the TASEP on a circle. Let $0 \leq p \leq 1$.
At each time $t \in \N_0$, given a particle configuration
$\eta_t=\{\eta_t(i)\}_{i \in \Lambda} \in \{0, 1 \}^{\Lambda}$,
let $A_t=\{(i,i+1): 1 \leq i \leq K,
\, \mbox{s.t.} \, \eta_t(i)=1, \eta_t(i+1)=0\}$,
where the periodicity 
$\eta(i+K)=\eta(i), i \in \Lambda$ is assumed and
the nearest-neighbor pair of sites $(K,K+1)$ 
is identified with $(K,1)$.
Every particle at site $i$ such that $(i,i+1) \in A_t$
has chance to move to its next site $i+1$,
since the site $i+1$ is vacant; $\eta_t(i+1)=0$.
But, in general, only a part of such particles move
depending on the parameter $p$ as follows.
We choose a subset of $A_t$ randomly, in the sense that
each pair of nearest-neighbor sites 
$(i,i+1) \in A_t$ is chosen independently
with probability $p$. The obtained subset of $A_t$
is written as $\widehat{A_t}$.
In the present paper, the total number of elements
included in a set $B$ is denoted by $|B|$
and, for $B \subset C$, the complementary set of $B$
in the set $C$ is expressed by $C \setminus B$.
(By definition $|C \setminus B|=|C|-|B|$.)
Then the probability that $\widehat{A_t}$ is chosen from $A_t$
is given by $p^{|\widehat{A_t}|} (1-p)^{|A_t|-|\widehat{A_t}|}$.
Only the particles at sites $\{i\}$ such that 
$(i,i+1) \in \widehat{A_t}$ indeed move to their next sites.
That is, the particle configuration at time $t+1$,
$\eta_{t+1}=\{\eta_{t+1}(i)\}_{i \in \Lambda}$,
is given by
\begin{equation}
\eta_{t+1}(i)=\left\{ \begin{array}{ll}
\eta_t(i)-1, & \quad \mbox{if $(i,i+1) \in \widehat{A_t}$},
\\
\eta_t(i)+1, & \quad \mbox{if $(i-1,i) \in \widehat{A_t}$},
\\
\eta_t(i), & \quad \mbox{otherwise}.
\end{array} \right.
\label{eqn:eta1}
\end{equation}
Here note that by definition of $\widehat{A_t}$,
if $(i,i+1) \in \widehat{A_t}$,
then $(i-1, i) \notin \widehat{A_t}$.
The parameter $p$ is called the {\it hopping probability}
and the above procedure is said to be 
the {\it parallel update rule}.
The total number of particles is conserved in the process,
which we write $L$ in the present paper.
We assume $1 \leq L \leq K$.

Given a particle configuration $\eta \in \{0,1\}^{\Lambda}$,
let $i_1=\min\{1 \leq i \leq K: \eta(i)=1\}$
and define
\begin{equation}
i_{j+1}=\min\{i_j < i \leq K: \eta(i)=1\},
\quad 1 \leq j \leq L-1,
\label{eqn:ij1}
\end{equation}
{\it i.e.}, $i_j$ is the site occupied by
the $j$-th particle, $1 \leq j \leq L$.
Then we put
\begin{equation}
n(j)=i_{j+1}-i_j-1, \quad 1 \leq j \leq L,
\label{eqn:n1}
\end{equation}
where $i_{L+1} \equiv i_1+K$.
That is, $n(j)$ gives the number of vacant sites
between the $j$-th and the $(j+1)$-th particles.
By (\ref{eqn:n1}) with (\ref{eqn:ij1}),
a configuration of vacancies $\n=\{n(j)\}_{j=1}^{L}$
is uniquely determined from the particle configuration 
$\eta=\{\eta(i)\}_{i=1}^K$.

We should note that $\n$ does not determine
$\eta$ uniquely, however, since the information on the position
of the first particle, $i_1$, is missing
in the map $\eta \to \n$.
This information may be, however, not important,
since here we consider the TASEP on a circle.
The stochastic process
$\n_t=\{n_t(j)\}_{j=1}^{L}, t \in \N_0$, 
obtained from $\eta_t, t \in \N_0$
by this map, is a special case of the ZRP
\cite{Eva00,EH05}.
As a consequence of general theory of ZRP
\cite{Eva97,EH05,KNT06,Kan07},
the probability distribution function in the stationary state
$\P_{L,N}(\n)$ of configuration $\n$
of vacancies is uniquely determined as follows.
Since the lattice size $K$ and the total number of
particles $L$ are conserved, the total number of
vacant sites $N \equiv \sum_{j=1}^L n(j)=K-L$
is also a constant.
We fix $1 \leq L, N \leq K$.
Then the configuration space of $\n$ is given by
\begin{equation}
\Omega_{L,N}=\left\{
\n=\{n(j)\}_{j=1}^{L} \in \{0,1, \dots, N\}^{L} :
\sum_{j=1}^{L} n(j)=N \right\}
\label{eqn:Omega}
\end{equation}
and
\begin{equation}
\P_{L,N}(\n)=\frac{1}{Z_{L,N}} \prod_{j=1}^{L} f(n(j)),
\quad \n=\{n(j) \}_{j=1}^{L} \in \Omega_{L,N},
\label{eqn:PLN}
\end{equation}
where \cite{Remark1}
\begin{equation}
f(n)=\left\{ \begin{array}{ll}
1, &\quad \mbox{if $n=0$}, \\
(1-p)^{n-1}, & \quad \mbox{if $n \geq 1$},
\end{array} \right.
\label{eqn:fn}
\end{equation}
and the partition function is given by \cite{KNT06,Kan07}.
\begin{eqnarray}
Z_{L,N} &\equiv& \sum_{\n \in \Omega_{L,N}}
\prod_{j=1}^{L} f(n(j)) \nonumber\\
&=& (1-p)^{N-1} L 
F\left(1-L, 1-N; 2; \frac{1}{1-p} \right)
\nonumber\\
&=&\frac{(-p)^{L+N} L}{(1-p)^{L+1}}
F\left( L+1, N+1; 2; \frac{1}{1-p} \right)
\label{eqn:Z1}
\end{eqnarray}
with the Gauss hypergeometric function \cite{AS72}
\begin{equation}
F(\alpha, \beta; \gamma; z)
=\sum_{n=0}^{\infty} \frac{(\alpha)_n (\beta)_n}{(\gamma)_n}
\frac{z^n}{n!},
\label{eqn:GaussH}
\end{equation}
$(\alpha)_0=1, (\alpha)_n=\alpha(\alpha+1) \cdots (\alpha+n-1),
n \geq 1$.
Note that the last equality in (\ref{eqn:Z1})
is due to Kummer's transformation \cite{AS72}
$$
F(\alpha, \beta; \gamma; z)
=(1-z)^{\gamma-\alpha-\beta} 
F(\gamma-\alpha, \gamma-\beta; \gamma; z).
$$
We write the expectation with respect to 
the stationary distribution (\ref{eqn:PLN}) as
$\E_{L,N}[\, \cdot \,]$ in this paper.

\begin{figure}
\includegraphics[width=0.7\linewidth]{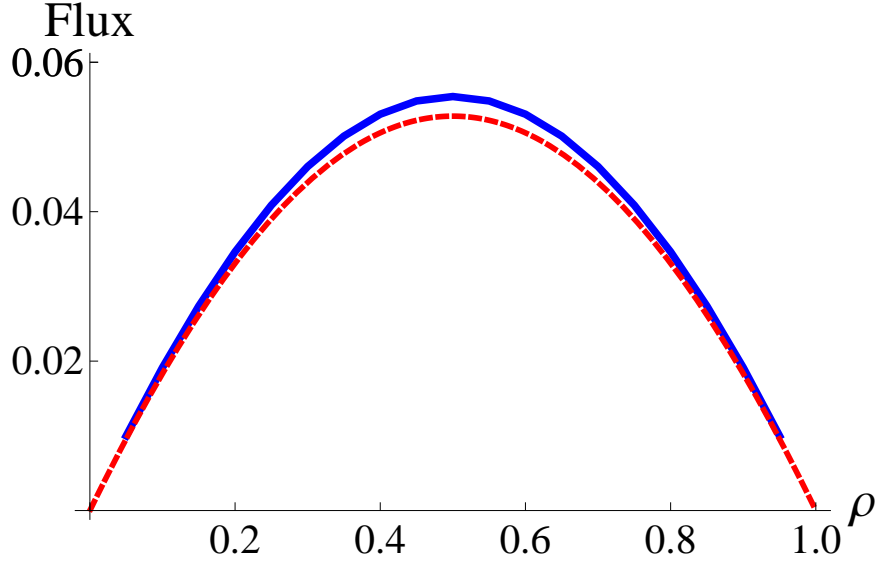}
\caption{
The fundamental diagrams.
The flux $J_{L,N}$ is plotted versus $\rho$ for $K=20$
with $p=0.2$ by a solid line.
The broken line shows the thermodynamic limit
(\ref{eqn:limit1}) with $p=0.2$.
We find that the difference between the results
for finite systems and the thermodynamic limit
is very small, in particular, when $\rho \simeq 0$,
$\rho \simeq 1.0$.
}
\label{fig:Fig1}
\end{figure}

For the $j$-th particle, $1 \leq j \leq L$,
if $n(j) \geq 1$, that is, if $n(j) \not=0$,
then that particle can move to the next site
with probability $p$ in a time-step.
Then if the velocity of $j$-th particle
is denoted by $V_j$,
the expectation of this random variable
in the stationary distribution $\P_{L,N}$
is given by
\begin{eqnarray}
\E_{L,N}[V_j]
&=& p \E_{L,N} [ \1(n(j) \geq 1) ]
\nonumber\\
&=& p \sum_{\n \in \Omega_{L,N}}
\1(n(j) \geq 1) \P_{L,N}(\n),
\label{eqn:EV1}
\end{eqnarray}
where $\1(\omega)$ is an indicator of an event $\omega$;
$\1(\omega)=1$ if $\omega$ occurs,
$\1(\omega)=0$ otherwise.
The average velocity (\ref{eqn:EV1}) is independent of $j$,
since the system is homogeneous in space, 
and it has been explicitly calculated as \cite{KNT06,Kan07}
\begin{eqnarray}
\E_{L,N}[V]
&=& \frac{\sum_{n=0}^{N-1} (-1)^{N+1-n}Z_{L,n}}
{Z_{L,N}}
\nonumber\\
&=& \frac{p}{L}
\frac{F(1-L,1-N;1;1/(1-p))}
{F(1-L,1-N;2;1/(1-p))}
\nonumber\\
&=& -\frac{1-p}{L} \frac{F(L,N; 1; 1/(1-p))}
{F(L+1, N+1; 2; 1/(1-p))}.
\label{eqn:EV2}
\end{eqnarray}
The density of particles is given by 
\begin{equation}
\rho=\frac{L}{L+N}=\frac{L}{K},
\label{eqn:rho}
\end{equation}
and the flux $J_{L, N}$ is defined by
\begin{equation}
J_{L,N}= \rho \E_{L, N}[V].
\label{eqn:J}
\end{equation}
If we plot $J_{L,N}$ 
versus $\rho$, we obtain a {\it fundamental diagram}
as demonstrated by Kanai \cite{Kan07}
(see Fig.\ref{fig:Fig1} in the present paper).
Moreover, Kanai {\it et al.}\cite{KNT06}
determined the {\it thermodynamic limit},
{\it i.e.} the scaling limit of $L \to \infty, N \to \infty$
with keeping $\rho=L/(L+N)$ be a constant,
for the average velocity and obtained
(see Fig.\ref{fig:Fig1}) 
\begin{equation}
\lim_{\substack{L \to \infty, N \to \infty;\\
\rho={\rm const.}}} \E_{L,N}[V]
=\frac{1-\sqrt{1-4 p \rho(1-\rho)}}{2 \rho},
\quad 0 \leq \rho, p \leq 1.
\label{eqn:limit1}
\end{equation}
This result coincides with the exact solution 
for an infinite system obtained
by Schadschneider and Schreckenberg \cite{SS93}.

In the present paper, we study velocity correlation functions
and show that, as extensions of the formula (\ref{eqn:EV2}),
they are generally expressed by using the Gauss hypergeometric
functions.
In particular, the covariance of velocities
$V$ and $V'$ of two particles at different sites,
\begin{equation}
\Cov_{L,N}[V,V']
= \E_{L,N}[V V']-\E_{L,N}[V]^2,
\label{eqn:Cov1}
\end{equation}
is studied in detail and it is shown that
\begin{equation}
\lim_{\substack{L \to \infty, N \to \infty; \\
\rho={\rm const.}}}
\Cov_{L,N}[V, V']=0.
\label{eqn:limit2}
\end{equation}
That is, velocities of particles are correlated
in the stationary state in any finite systems,
but it is proved that they become independent 
asymptotically 
in the thermodynamic limit in the discrete-time
TASEP on a circle.

The paper is organized as follows.
In Sec.II.A velocity correlation functions
are defined and the general formula is derived.
As special cases, the obtained expressions of average
velocity and covariance of velocities of two particles
are studied in detail in Sec.II.B and C, respectively.
Sec.III is devoted to proving asymptotic
independence of velocities (\ref{eqn:limit2})
in the thermodynamic limit.
Concluding remarks are given in Sec.IV.
Appendix A is given for showing the derivation
of a Riccati equation, which governs 
the covariance of velocities and is used in the 
proof in Sec.III.

\section{Velocity Correlation Functions}
\subsection{General Formula}

Let $1 \leq k \leq L$.
From the $L$ particles, we pick up $k$ distinct particles
arbitrarily;
let $1 \leq j_1 < j_2 < \cdots < j_{k} \leq L$,
and consider a set of $k$ particles such that
the $\ell$-th particle in this set, $1 \leq \ell \leq k$, is
originally the $j_{\ell}$-th particle in the
whole particle systems.
Write the velocity of the $j_{\ell}$-th particle
by $V_{j_{\ell}}, 1 \leq \ell \leq k$.
The velocity correlation function for the $k$ particle
is then defined by
\begin{eqnarray}
F_{L,N}(\{V_{j_{\ell}}\}_{\ell=1}^k)
&=& \E_{L,N} \left[ \prod_{\ell=1}^{k} V_{j_{\ell}} \right]
\nonumber\\
&=& p^k \sum_{\n \in \Omega_{L,N}}
\prod_{\ell=1}^{k} 
\1(n(j_{\ell}) \geq 1) \P_{L,N}(\n).
\label{eqn:F1}
\end{eqnarray}
Since the stationary distribution function is given by
the factorized form (\ref{eqn:PLN}), it is written as
\begin{eqnarray}
&& F_{L,N}(\{V_{j_{\ell}}\}_{\ell=1}^{k})
=\frac{p^{k}}{Z_{L,N}}
\sum_{n(1)=0}^{N} \sum_{n(2)=0}^{N} \cdots
\sum_{n(L)=0}^{N} \prod_{\ell=1}^{k} \1(n(j_{\ell}) \geq 1)
\prod_{j=1}^{L} f(n(j)) 
\1 \left( \sum_{j=1}^{L} n(j)=N \right)
\nonumber\\
&& \quad =
\frac{p^k}{Z_{L,N}} \prod_{j \in \I_L \setminus \J_k}
\sum_{n(j)=0}^N f(n(j))
\prod_{\ell \in \J_k} \sum_{n(\ell)=1}^N f(n(\ell))
\1 \left( \sum_{i=1}^{L} n(i)=N \right)
\nonumber\\
&& \quad = \frac{p^k}{Z_{L,N}}
\prod_{j \in \I_L \setminus \J_k} \sum_{n(j)=0}^N f(n(j))
\prod_{\ell \in \J_k} \left( \sum_{n(\ell)=0}^N f(n(\ell))-f(0) \right)
\1 \left( \sum_{i=1}^L n(i)=N \right),
\label{eqn:FLNB1}
\end{eqnarray}
where $\I_L=\{1,2, \dots, L\}$ and
$\J_k=\{j_1, j_2, \dots, j_{k}\}$.

We perform the binomial expansion
\begin{equation}
\prod_{\ell \in \J_k} \left( \sum_{n(\ell)=0}^N f(n(\ell))-f(0) \right)
=\sum_{\K \subset \J_k}
(-f(0))^{|\J_k \setminus \K|}
\prod_{\ell \in \K} \sum_{n(\ell)=0}^N f(n(\ell)),
\label{eqn:expansion}
\end{equation}
where the first summation in the RHS is taken over all subsets $\K$
of $\J_k$ and $|\J_k \setminus \K|$ is the number of
elements in the complementary set of $\K$ in $\J_k$.
Note that $\I_L=(\I_L \setminus \J_k) \cup \K \cup (\J_k \setminus \K)$
and $i \in \J_k \setminus \K$ implies $n(i)=0$,
since the weight $f(0)/Z_{L,N}$ is assigned in the expansion.
Therefore we can set
$\sum_{i=1}^{L} n(i)=\sum_{i \in (\I_L \setminus \J_k) \cup \K} n(i)$
in (\ref{eqn:FLNB1}).
Since we set $f(0)=1$ as (\ref{eqn:fn}), we obtain
\begin{eqnarray}
&& F_{L, N}(\{V_{j_{\ell}}\}_{\ell=1}^k)
\nonumber\\
&& = \frac{p^k}{Z_{L,N}} \sum_{\K \in \J_k}
(-1)^{|\J_k \setminus \K|}
\prod_{j \in \I_L \setminus \J_k}
\sum_{n(j)=0}^N f(n(j)) \prod_{\ell \in \K} \sum_{n(\ell)=0}^N f(n(\ell))
\1 \left( \sum_{i \in (\I_L \setminus \J_k) \cup \K}
n(i)=N \right)
\nonumber\\
&& = \frac{p^k}{Z_{L,N}} \sum_{\K \in \J_k}
(-1)^{|\J_k \setminus \K|}
\prod_{i \in (\I_L \setminus \J_k) \cup \K}
\sum_{n(i)=0}^N f(n(i))
\1 \left( \sum_{i \in (\I_L \setminus \J_k) \cup \K}
n(i)=N \right).
\label{eqn:F2}
\end{eqnarray}
When $|\J_k \setminus \K|=s, 0 \leq s \leq k$, 
$|\K|=k-s$ and thus
$|(\I_L \setminus \J_k) \cup \K|
=|\I_L \setminus \J_k|+|\K|
=(L-k)+(k-s)=L-s$.
Therefore
\begin{eqnarray}
&& \prod_{i \in (\I_L \setminus \J_k) \cup \K}
\sum_{n(i)=0}^{N} f(n(i)) 
\1 \left( \sum_{i \in (\I_L \setminus \J_k) \cup \K)}
n(i)=N \right)
\nonumber\\
&& \quad = \prod_{p=1}^{L-s} 
\sum_{n(p)=0}^N f(n(p)) 
\1 \left( \sum_{p=1}^{L-s} n(p)=N \right)
\nonumber\\
&& \quad = \sum_{\n \in \Omega_{L-s, N}}
\prod_{p=1}^{L-s} f(n(p))=Z_{L-s, N}.
\nonumber
\end{eqnarray}
Since the number of distinct subsets $\K$ in $\J_k$
satisfying $|\J_k \setminus \K|=s$ is
$\displaystyle{ {k \choose s} }, 0 \leq s \leq k$,
(\ref{eqn:F2}) is equal to 
$\{p^k/Z_{L,N}\}
\sum_{s=0}^{k}(-1)^s {k \choose s} Z_{L-s,N}$.
The result does not
depend on the choice of particle positions
$\J_{k}=\{j_1, j_2, \dots, j_k\}$,
but depends only on the total number $k$
of particles, whose velocity correlation is
calculated.
This special property comes from the factorized
form (\ref{eqn:PLN}) of the stationary distribution
in the present system, in which the factor $f(n)$
are independent of the system sizes, $L$ and $N$, 
as given by (\ref{eqn:fn}).
We summarize the result by the following formula,
\begin{eqnarray}
F_{L,N}(k) &\equiv&
F_{L,N}(\{V_{j_{\ell}}\}_{\ell=1}^k)
\nonumber\\
&=& \frac{p^k}{Z_{L,N}}
\sum_{s=0}^k (-1)^s {k \choose s}
Z_{L-s, N}
\nonumber\\
&=& \frac{p^k}{L F(1-L, 1-N; 2; 1/(1-p))}
\nonumber\\
&& \quad \times
\sum_{s=0}^k (-1)^s (L-s) 
{k \choose s} F \left(1-L+s,1-N;2; \frac{1}{1-p} \right)
\nonumber\\
&=&
\frac{p^k}{L F(L+1, N+1; 2; 1/(1-p))}
\nonumber\\
&& \quad \times
\sum_{s=0}^k (L-s) {k \choose s} 
\left(\frac{1-p}{p} \right)^s
F \left( L-s+1, N+1; 2; \frac{1}{1-p} \right).
\label{eqn:main1}
\end{eqnarray}
It should be noted that still velocities are correlated
in the sense that 
$F_{L,N}(k) \not=(F_{L,N}(1))^k, k \geq 2$.
In other words,
$$
\E_{L,N} \left[ \prod_{\ell=1}^k V_{j_{\ell}} \right]
\not= \prod_{\ell=1}^k \E_{L,N} \left[ V_{j_{\ell}} \right],
\quad k \geq 2.
$$

\subsection{Average Velocity}

By setting $k=1$ in the general formula (\ref{eqn:main1}),
we obtain
\begin{eqnarray}
&& \E_{L,N}[V] = F_{L,N}(1) 
= \frac{p}{Z_{L,N}}(Z_{L,N}-Z_{L-1,N})
\nonumber\\
&& = p \frac{L F(1-L,1-N;2;1/(1-p))
-(L-1)F(2-L,1-N;2;1/(1-p))}
{L F(1-L,1-N;2;1/(1-p))}
\nonumber\\
&& =
\frac{L p F(L+1, N+1; 2; 1/(1-p))
+(L-1)(1-p) F(L, N+1; 2; 1/(1-p))}
{L F(L+1, N+1; 2; 1/(1-p))}.
\label{eqn:EV3}
\end{eqnarray}

Now we show that the last expression of (\ref{eqn:EV3}) is equal to
the last expression of (\ref{eqn:EV2}).
First we rewrite the numerator of (\ref{eqn:EV3}) as
\begin{eqnarray}
&& -(1-p) \Bigg[L F \left(L+1,N+1; 2; \frac{1}{1-p} \right)
-\frac{L}{1-p} F \left(L+1,N+1;2; \frac{1}{1-p} \right)  
\nonumber\\
&& \qquad \qquad -(L-1) F \left(L,N+1; 2; \frac{1}{1-p} \right) \Bigg].
\label{eqn:EV3N}
\end{eqnarray}
If we use the recurrence relation of the Gauss hypergeometric series
\cite{AS72}
$$
\alpha z F(\alpha+1,\beta+1;\gamma+1;z)=
\gamma \Big\{F(\alpha,\beta+1;\gamma;z)
-F(\alpha,\beta;\gamma;z) \Big\}
$$
for the second term in (\ref{eqn:EV3N}), 
(\ref{eqn:EV3N}) becomes
\begin{eqnarray}
&&
-(1-p) \Bigg[ L F \left(L+1,N+1; 2; \frac{1}{1-p}\right)
-F \left(L,N+1; 1; \frac{1}{1-p} \right)
\nonumber\\
&& \qquad \qquad 
+F \left(L,N;1; \frac{1}{1-p} \right)
-(L-1) F \left(L,N+1;2; \frac{1}{1-p} \right)
 \Bigg].
\label{eqn:EV4}
\end{eqnarray}
Next we apply the formula \cite{AS72}
$$
(\gamma-\alpha-1) F(\alpha,\beta; \gamma; z)
+\alpha F(\alpha+1,\beta; \gamma;z)
=(\gamma-1) F(\alpha,\beta;\gamma-1;z)
$$
to the first and the fourth terms in (\ref{eqn:EV4}).
Then the sum of them becomes
$$
L F \left(L+1,N+1; 2; \frac{1}{1-p} \right)
-(L-1) F \left(L,N+1; 2; \frac{1}{1-p} \right)
=F \left(L,N+1; 1; \frac{1}{1-p} \right),
$$
which is cancelled by the second term in (\ref{eqn:EV4}).
Therefore, the numerator of (\ref{eqn:EV3}) is equal to
$-(1-p) F (L,N;1; 1/(1-p))$, 
and the equivalence of the last expression of (\ref{eqn:EV3}) and
the last expression of (\ref{eqn:EV2}) is confirmed.

\subsection{Covariance of Velocity}

By the first expression in (\ref{eqn:main1}) for $k=2$, we obtain
\begin{eqnarray}
\E_{L,N}[VV']
&=& F_{L, N}(2)
\nonumber\\
&=& \frac{p^2}{Z_{L,N}}
(Z_{L,N}-2 Z_{L-1, N}+Z_{L-2,N}).
\label{eqn:EVV1}
\end{eqnarray}
Then the covariance of velocities (\ref{eqn:Cov1})
is given by
\begin{equation}
\Cov_{L,N}[V, V']
=p^2 \left\{ \frac{Z_{L-2,N}}{Z_{L,N}}
-\left( \frac{Z_{L-1,N}}{Z_{L,N}} \right)^2 \right\}.
\label{eqn:Cov2}
\end{equation}
By the expression of partition function (\ref{eqn:Z1}) 
using the hypergeometric
function, it is written as
\begin{eqnarray}
\Cov_{L,N}[V,V']
&=& p^2 \left\{
\frac{(L-2) F(3-L, 1-N; 2; 1/(1-p))}
{L F(1-L, 1-N; 2; 1/(1-p))} \right.
\nonumber\\
&& \qquad \qquad \left.
-\left( \frac{(L-1) F(2-L,1-N;2;1/(1-p))}
{L F(1-L,1-N;2;1/(1-p))} \right)^2 \right\}
\nonumber\\
&=& (1-p)^2 \left\{
\frac{(L-2) F(L-1,N+1;2; 1/(1-p))}
{L F(L+1, N+1; 2; 1/(1-p))} \right.
\nonumber\\
&& \qquad \qquad \left.
-\left( \frac{(L-1) F(L,N+1; 2; 1/(1-p))}
{L F(L+1, N+1; 2; 1/(1-p))} \right)^2 \right\}.
\label{eqn:Cov3}
\end{eqnarray}
If we use the recurrence relation of the
Gauss hypergeometric function \cite{AS72}
\begin{eqnarray}
&& \alpha(1-z) F(\alpha+1,\beta; \gamma; z)
+ \Big[\gamma-2\alpha+(\alpha-\beta)z \Big] F(\alpha,\beta; \gamma; z)
\nonumber\\
&& \qquad \qquad 
+(\alpha-\gamma) F(\alpha-1,\beta; \gamma; z)=0,
\nonumber
\end{eqnarray}
the second expression of 
(\ref{eqn:Cov3}) is rewritten as
\begin{eqnarray}
\Cov_{L,N}[V, V'] &=& (1-p)^2
\left[ \frac{1}{1-p}-1
+\left(1-\frac{L-N-1}{2(L-1)} \frac{1}{1-p} \right)^{2}
\right. \nonumber\\
&& \qquad \qquad \left.
-\left\{Y_{L,N}(p)-\left(1-\frac{L-N-1}{2(L-1)}
\frac{1}{1-p} \right)\right\}^2  \right],
\label{eqn:Cov4}
\end{eqnarray}
where 
\begin{equation}
Y_{L,N}(p)
\equiv \frac{(L-1) F (L,N+1; 2; 1/(1-p))}
{L F(L+1,N+1;2; 1/(1-p))}.
\label{eqn:Y1}
\end{equation}

\begin{figure}
\includegraphics[width=0.7\linewidth]{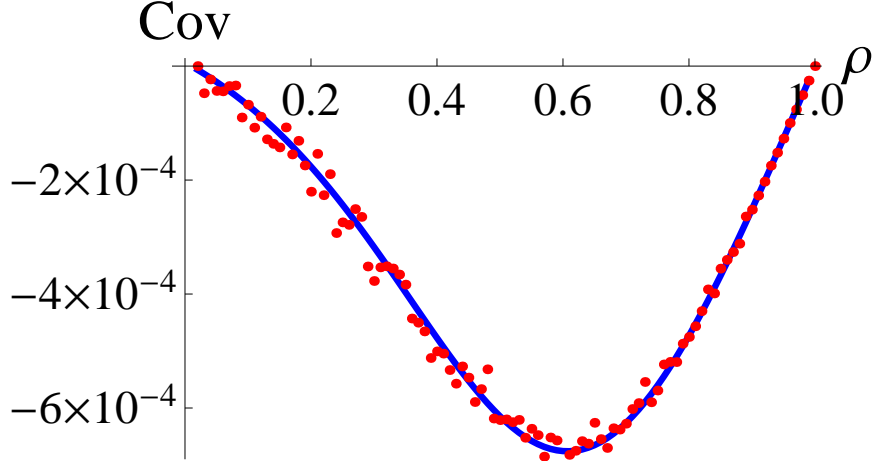}
\caption{
The exact solution of the covariance of velocities
(\ref{eqn:Cov3}) is plotted versus the particle density $\rho$
for $K=100$ with $p=0.5$.
Computer simulations are also performed and numerical
results are dotted.
}
\label{fig:Fig2}
\end{figure}

Figure \ref{fig:Fig2} shows $\Cov_{L,N}[V, V']$ given by
(\ref{eqn:Cov3}) for $K=100$ with $p=0.5$
as a function of the density of particles $\rho$.
This figure shows that velocities are negatively correlated.
We performed computer simulations
for a variety of systems with $K=100$ and $p=0.5$
by changing the number of particles $1 \leq L \leq K$.
For each trial, we discarded first $2 \times 10^3$ time-step data
and used $8 \times 10^3$ time-step data for evaluating 
the covariance of velocities.
In Fig.\ref{fig:Fig2}, each dot indicates the averaged value of
$10^4$ trials.
We can find that the simulation data coincide very well
with the exact solution (\ref{eqn:Cov3}).

\section{Asymptotic Independence of Velocities}
\subsection{Riccati Equations}

Here we consider the average velocity (\ref{eqn:EV2})
as a function of the hopping probability $p$
and express it by
\begin{equation}
v_{L,N}(p)= \E_{L,N}[V].
\label{eqn:v1}
\end{equation}
Kanai {\it et al.} \cite{KNT06} proved that it solves
a Riccati equation
\begin{equation}
\frac{dv_{L,N}(p)}{dp}
+\frac{L}{p(1-p)} v_{L,N}(p)^2 
- \frac{L+N}{p(1-p)} v_{L,N}(p)
+\frac{N}{1-p} =0.
\label{eqn:Riccati1}
\end{equation}
From this equation, 
the thermodynamic limit (\ref{eqn:limit1}) was concluded.
(See also (\ref{eqn:v0}) below.)

We have found that $Y_{L,N}(p)$
defined by (\ref{eqn:Y1}) also solves a Riccati equation
in the form
\begin{eqnarray}
&& \frac{d Y_{L,N}(p)}{dp}
+\frac{L-1}{p}\frac{v_{L-1,N}(p)}{v_{L,N}(p)} Y_{L,N}(p)^2
\nonumber\\
&& \qquad 
+ \left[
\frac{d}{dp} \log \frac{v_{L-1,N}(p)}{v_{L,N}(p)} 
+\frac{ 2(L-1)p-(L+N-1)}{p(1-p)}
\right] Y_{L,N}(p)
\nonumber\\
&& \qquad \qquad \qquad
-\frac{L-1}{1-p} \frac{v_{L,N}(p)}{v_{L-1,N}(p)}=0.
\label{eqn:Riccati2}
\end{eqnarray}
The derivation is given in Appendix A.
Note that by (\ref{eqn:Riccati1}) we obtain
the equation
\begin{eqnarray}
&& \frac{d}{dp} \log \frac{v_{L-1,N}(p)}{v_{L,N}(p)} 
=\frac{d v_{L-1,N}(p)/dp}{v_{L-1,N}(p)}
-\frac{d v_{L,N}(p)/dp}{v_{L,N}(p)}
\nonumber\\
&&
= -\frac{1}{p(1-p)}
\left[ (L-1) v_{L-1,N}(p)-Lv_{L,N}(p)+1
+N p \left( \frac{1}{v_{L-1,N}(p)}-\frac{1}{v_{L,N}(p)} \right) 
\right].
\label{eqn:Riccati2b}
\end{eqnarray}
That is, the Riccati equation (\ref{eqn:Riccati2}) for $Y_{L,N}(p)$,
which governs $\Cov_{L,N}[V,V']$
through (\ref{eqn:Cov4}), is coupled with the Riccati equations
(\ref{eqn:Riccati1}) for $v_{L,N}(p)$ and $v_{L-1,N}(p)$.

\subsection{Large-size expansion and thermodynamic limit}

Following the procedure given by \cite{KNT06},
we consider the power expansion of the quantities
with respect to the inverse of system size, $1/K$
with $K=L+N$,
\begin{eqnarray}
\label{eqn:exp1}
&& v_{L,N}(p)
=v_0+v_1 \frac{1}{K}+v_2 \frac{1}{K^2}+ \cdots,
\\
\label{eqn:exp2}
&& v_{L-1,N}(p)
=v_0'+v_1' \frac{1}{K}+v_2' \frac{1}{K^2}+ \cdots,
\\
\label{eqn:exp3}
&& Y_{L,N}(p)=Y_0+Y_1 \frac{1}{K}+Y_2 \frac{1}{K^2}+ \cdots,
\end{eqnarray}
where the coefficients $v_i, v_i', Y_i, i=0,1,2, \dots$
are assumed to be functions of $p$ and $\rho$.
Putting (\ref{eqn:exp1}) and (\ref{eqn:exp2})
into (\ref{eqn:Riccati1}) and its modification obtained by setting
$L \to L-1$, and taking the thermodynamic
limit $K \to \infty$ with $\rho=L/K=$ const.,
the first terms in (\ref{eqn:exp1}) and (\ref{eqn:exp2})
are determined as \cite{KNT06}
\begin{equation}
v_0=v_0'=\frac{1-\sqrt{1-4p \rho(1-\rho)}}{2 \rho}.
\label{eqn:v0}
\end{equation}
This result implies (\ref{eqn:limit1}).

Similarly, we put (\ref{eqn:exp1})-(\ref{eqn:exp3})
into (\ref{eqn:Riccati2}) with (\ref{eqn:Riccati2b}).
In the thermodynamic limit, the differential equation
is reduced to the algebraic equation
$$
\rho(1-p) Y_0^2-(1-2 p \rho) Y_0-\rho p=0
$$
for $Y_0$, which is solved as
\begin{equation}
Y_0 =
\lim_{\substack{K \to \infty; \\ \rho={\rm const.}}}
Y_{L,N}(p)
= \frac{(1-2\rho p) \pm \sqrt{1-4\rho p(1-\rho)}}{2\rho(1-p)}.
\label{eqn:Y0}
\end{equation}

On the other hand, in the similar way, we can show that
(\ref{eqn:Cov4}) gives
\begin{equation}
\lim_{\substack{K \to \infty; \\ \rho={\rm const.}}} 
\Cov_{L,N}[V, V'] = 
p(1-p)+\left[\frac{1-2\rho p}{2\rho} \right]^2 
-\left[(1-p)Y_{0}-\left( \frac{1-2\rho p}{2\rho} \right ) \right ]^2.
\label{eqn:CovB1}
\end{equation}
If we apply the result (\ref{eqn:Y0}), (\ref{eqn:CovB1}) turns to be
\begin{eqnarray}
\lim_{\substack{K \to \infty;\\ \rho={\rm const.}}}
\Cov_{L,N}[V,V'] &=& p(1-p)
+\left[\frac{1-2\rho p}{2\rho} \right]^2
-\frac{1-4\rho p(1-\rho)}{4\rho^2} \nonumber \\
&=& 0.
\label{eqn:CovB2}
\end{eqnarray}
Vanishing of the covariance implies that
velocities of particles of the discrete-time TASEP
become independent asymptotically
in the thermodynamic limit in the stationary state
on a circle.

\begin{figure}
\includegraphics[width=0.7\linewidth]{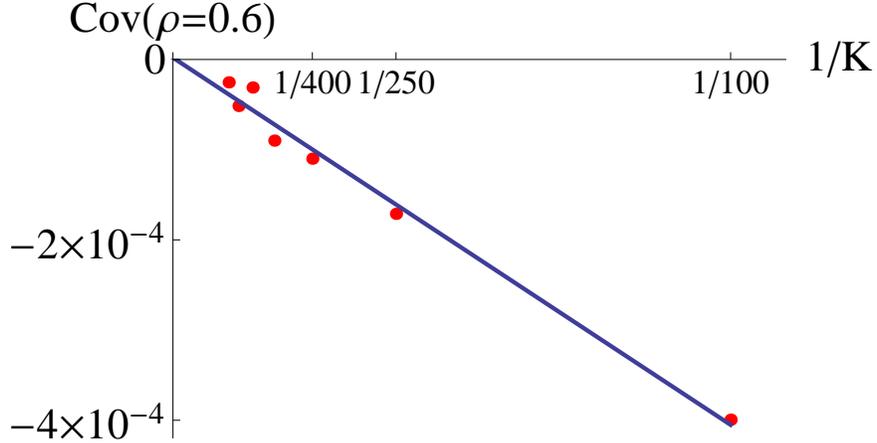}
\caption{Numerical fitting of $\Cov_{L,N}[V,V']$
evaluated by computer simulations versus $1/K$
for $p=0.5, \rho=0.6$.
The data shows that $\Cov_{L,N}[V,V']$ becomes zero
as $K=L+N \to \infty$ in the form (\ref{eqn:fitting1})
with $c(p=0.5, \rho=0.6) \simeq -4 \times 10^{-2}$.
In each trial of computer simulation,
we discarded first $2 \times 10^4$ time-step data
and used $10^4$ time-step data for evaluating
the covariance of velocities and each dot
in this figure indicates the averaged value 
over $10^{4}$ trials.
}
\label{fig:Fig3}
\end{figure}

Figure \ref{fig:Fig3} shows the numerical data demonstrating
$\Cov_{L,N}[V,V'] \to 0$ as $K=L+N \to \infty$.
Here we set $p=0.5$ and $\rho=0.6$ and performed
computer simulations by increasing $K$ from 100 to 1000
by 150, in which $\Cov_{L,N}[V,V']$ with
$L=\rho K$ and $N=(1-\rho)K$ are evaluated.
The linear fitting of $\Cov_{L,N}[V,V']$ versus $1/K$
gives
\begin{equation}
\Cov_{L,N}[V,V'] \simeq \frac{c(p,\rho)}{K}
\label{eqn:fitting1}
\end{equation}
with
$c(p=0.5, \rho=0.6) \simeq - 4 \times 10^{-2}$.

\section{Concluding Remarks}

In the present paper, we study a discrete-time version
of the TASEP on a circle developed by the parallel update rule.
This version is easily simulated by a computer.
As a matter of fact, we have checked the validity
of our exact solutions for finite-size systems
by comparing them with the numerical simulation data
as shown by Fig.\ref{fig:Fig2}.

From the viewpoint of statistical mechanics,
it is important to discuss the thermodynamic limit
for non-equilibrium steady states realized in
the present model.
It is rather difficult, however, to define
parallel update dynamics for a system
with an infinite number of particles, 
since the number of updated sites
can be infinity, {\it i.e.}, $|\widehat{A_t}|=\infty$
using the notation in Sec.I.
Kanai {\it et al.}\cite{KNT06} overcame this difficulty
and determined the thermodynamic limit of 
average velocity (\ref{eqn:limit1}).
They obtained the differential
equation which governs the average velocity
and took the thermodynamic limit in the equation.
In the present paper, we extend their procedure
for the covariance of two-particle velocities
and showed that it becomes asymptotically zero
in the thermodynamic limit (\ref{eqn:CovB2}).

It should be remarked that
both equations which govern the average velocity
and the covariance are given by the Riccati-type
differential equations with respect to the
hopping probability $p$.
As a matter of course, the Riccati equation (\ref{eqn:Riccati2})
for $Y_{L,N}(p)$, which governs $\Cov_{L,N}[V,V']$
through (\ref{eqn:Cov4}), is coupled with 
the equations (\ref{eqn:Riccati1}) for the average velocities
(see (\ref{eqn:Riccati2b})),
and thus it becomes much complicated.
Further study of hierarchy in the coupled system of differential equations
which determine the third and higher-order
moments of velocities will be an
interesting future problem.

\begin{acknowledgments}
The present authors would like to thank
M. Kanai for useful discussion on the problem.
This work is supported in part by
the Grant-in-Aid for Scientific Research (C)
(No.21540397) of Japan Society for
the Promotion of Science.
\end{acknowledgments}

\appendix
\section{Derivation of Riccati equation (\ref{eqn:Riccati2})}

By definitions (\ref{eqn:Y1}) and (\ref{eqn:v1}) with 
(\ref{eqn:EV2}),
\begin{equation}
\frac{v_{L-1,N}(p)}{v_{L,N}(p)} Y_{L,N}(p)
=\frac{F(L-1,N;1,z)}{F(L,N;1,z)},
\label{eqn:A1}
\end{equation}
where we put
\begin{equation}
z=\frac{1}{1-p}.
\label{eqn:z1}
\end{equation}
By the following formula of the Gauss hypergeometric function
\cite{AS72}
$$
\frac{d}{dz} \left[ 
\frac{(1-z)^{L+N-1}}{z^{L-1}}
F(L,N; 1; z) \right]
=(1-L) \frac{(1-z)^{L+N-2}}{z^L} F(L-1,N; 1; z),
$$
(\ref{eqn:A1}) is written as
\begin{equation}
\frac{v_{L-1,N}(p)}{v_{L,N}(p)} Y_{L,N}(p)
=\frac{z(1-z)}{1-L}
\frac{1}{w(z)} \frac{dw(z)}{dz}
\label{eqn:A4}
\end{equation}
with
\begin{equation}
w(z) =\frac{(1-z)^{L+N-1}}{z^{L-1}}
F(L,N;1;z).
\label{eqn:w1}
\end{equation}

Here we consider the generalized hypergeometric 
differential equation
\begin{eqnarray}
&& \frac{d u(z)}{dz^2}
+\sum_{i=1}^3 \frac{1-\lambda_i-\lambda_i'}{z-a_i}
\frac{du(z)}{dz}
\nonumber\\
&& \qquad
+ \sum_{i=1}^3 \frac{\lambda_i \lambda_i'}{z-a_i}
\prod_{1 \leq j \leq 3; j \not=i}
(a_i-a_j)
\frac{u(z)}{(z-a_1)(z-a_2)(z-a_3)}=0,
\label{eqn:AG1}
\end{eqnarray}
where Fuchs' relation
$\sum_{i=1}^3(\lambda_i+\lambda_i')=1$ is 
assumed to be satisfied.
The solution of (\ref{eqn:AG1}) is expressed by
\begin{equation}
u(z)=P \left\{ \begin{array}{cccc}
a_1 & a_2 & a_3 & \cr
\lambda_1 & \lambda_2 & \lambda_3 & z \cr
\lambda_1' & \lambda_2' & \lambda_3' & 
\end{array} \right\},
\label{eqn:AG2}
\end{equation}
which is called Riemann's $P$ function \cite{AS72}.
As a special case, the Gauss hypergeometric function (\ref{eqn:GaussH})
is given by
\begin{equation}
F(\alpha,\beta; \gamma; z)
=P \left\{ \begin{array}{cccc}
0 & 1 & \infty & \cr
0 & 0 & \alpha & z \cr
1-\gamma & \gamma-\alpha-\beta & \beta & 
\end{array} \right\}.
\label{eqn:AG3}
\end{equation}
In general, the following relations holds,
\begin{eqnarray}
\label{eqn:AG4}
P \left\{ \begin{array}{cccc}
a_1 & a_2 & a_3 & \cr
\lambda_1 & \lambda_2 & \lambda_3 & z \cr
\lambda_1' & \lambda_2' & \lambda_3' & 
\end{array} \right\}
&=& P \left\{ \begin{array}{cccc}
a_2 & a_1 & a_3 & \cr
\lambda_2 & \lambda_1 & \lambda_3 & z \cr
\lambda_2' & \lambda_1' & \lambda_3' & 
\end{array} \right\}, \\
\label{eqn:AG5}
(z-a_1)^k
P \left\{ \begin{array}{cccc}
a_1 & a_2 & \infty & \cr
\lambda_1 & \lambda_2 & \lambda_3 & z \cr
\lambda_1' & \lambda_2' & \lambda_3' & 
\end{array} \right\}
&=& P \left\{ \begin{array}{cccc}
a_1 & a_2 & \infty & \cr
\lambda_1+k & \lambda_2 & \lambda_3-k & z \cr
\lambda_1'+k & \lambda_2' & \lambda_3'-k & 
\end{array} \right\}.
\end{eqnarray}
By (\ref{eqn:AG3}), (\ref{eqn:w1}) is written as
$$
w(z) = (-1)^{L+N-1} (z-1)^{L+N-1} z^{1-L} 
P \left\{ \begin{array}{cccc}
0 & 1 & \infty &\cr
0 & 0 & L & z \cr
0 & 1-L-N & N &
\end{array} \right\}.
$$
Using the relations (\ref{eqn:AG4}) and (\ref{eqn:AG5}),
we can show that
\begin{equation}
w(z)= P \left\{ \begin{array}{cccc}
0 & 1 & \infty & \cr
1-L & L+N-1 & L-N & z \cr
1-L & 0 & 0 & 
\end{array} \right\}.
\label{eqn:w2}
\end{equation}
It implies that $w(z)$ solves the differential equation
\begin{equation}
\frac{d^2 w(z)}{dz^2}
-\frac{(L-N+1)z-2L+1}{(1-z)z}
\frac{d w(z)}{dz}
+\frac{(L-1)^2}{(1-z)z^2} w(z) =0.
\label{eqn:A5}
\end{equation}
Since $dw(z)/dz$ is related with $Y_{L,N}(p)$
by (\ref{eqn:A4}), (\ref{eqn:A5})
gives the first-order differential equation
for $Y_{L,N}(p)$.
By straightforward calculation, 
(\ref{eqn:Riccati2}) is derived.




\begin{thebibliography}{}
\bibitem{Remark1}
If we set $f(n)=(1-p)^{n-1} f(1)^n/f(0)^{n-1}$ for $n \geq 1$
and $Z_{L,N}=\sum_{\n \in \Omega_{L,N}} \prod_{j=1}^{L} f(n(j))$,
the values of $f(0)$ and $f(1)$ can be arbitrarily chosen
without changing $\P_{L,N}(\n)$ given by (\ref{eqn:PLN}).
In \cite{KNT06}, they are chosen as
$f(0)=1-p, f(1)=(1-p)/p$.
Here we simply set $f(0)=f(1)=1$.
\end{thebibliography}
\end{document}